\documentclass[aps,prl,twocolumn,groupedaddress]{revtex4}

\newcommand {\GdFeO}{GdFeO$_3$}
\newcommand {\CaIrO}{CaIrO$_3$}
\newcommand {\SrIrO}{SrIrO$_3$}
\newcommand {\ZrTe}{ZrTe$_5$}
\newcommand {\HgCdTe}{Hg$_{1-x}$Cd$_{x}$Te}
\newcommand {\CdAs}{Cd$_3$As$_2$}
\newcommand {\jeff}{\it j$_{\rm{eff}}$}
\newcommand {\kF}{$k_{\rm{F}}$}
\newcommand {\EF}{$E_{\rm F}$}
\newcommand {\vF}{$v_{\rm{F}}$}
\newcommand {\BQL}{$B _{\rm QL}$}
\newcommand {\Bth}{$B _{\rm th}$}

\newcommand {\DeltaExp}{$\varDelta_{\rm exp}$}
\newcommand {\DeltaCal}{$\varDelta_{\rm cal}$}
\newcommand {\DeltaEF}{$\varDelta E_{\rm F}$}
\newcommand {\Rhozz}{$\rho_{zz}$}
\newcommand {\Rhoxx}{$\rho_{xx}$}
\newcommand {\Rhoyx}{$\rho_{yx}$}

\newcommand {\nUnit}{$\rm{cm}^{-3}$}
\newcommand {\muUnit}{$\rm{cm}^{2}/\rm{Vs}$}
\newcommand {\BaIaz}{$B \parallel I \parallel a(z)$}
\newcommand {\BazIbx}{$B \parallel a(z), I \parallel b(x)$}

\newcommand {\Baz}{$B \parallel a(z)$}
\newcommand {\Bparaa}{$B \parallel a$}
\newcommand {\Bparac}{$B \parallel c$}
\newcommand {\BaIa}{$B \parallel I \parallel a$}
\newcommand {\BaIb}{$B \parallel a, I \parallel b$}
\newcommand {\Gzz}{$\sigma_{zz}$}
\newcommand {\Gxx}{$\sigma_{xx}$}
\newcommand {\Gzero}{$\sigma_{0}$}
\newcommand {\Gone}{$\sigma_{1}$}

\newcommand {\RThrOm}{$R_{zz}^{3 \omega}$}
\newcommand {\VThrOm}{$V_{zz}^{3 \omega}$}
\newcommand {\RThrOmNorm}{$R_{zz}^{3 \omega}/R_{zz}^{1 \omega}$}
\newcommand {\ROneOmDef}{$R_{zz}^{1 \omega} = V_{zz}^{1 \omega}/I$}
\newcommand {\RThrOmDef}{$R_{zz}^{3 \omega} = -V_{zz}^{3 \omega}/I$}

\usepackage{graphicx,bm}
\usepackage{pifont}
\usepackage{amsmath,amssymb}
\usepackage{color}

\begin{document}

\title{Field-induced multiple metal-insulator crossovers of correlated Dirac electrons of perovskite {\CaIrO}}

\author{R. Yamada$^{1}$, J. Fujioka$^{2}$, M. Kawamura$^{3}$, S. Sakai$^{3}$, M. Hirayama$^{1,3}$, R. Arita$^{1,3}$, T. Okawa$^{1}$, D. Hashizume$^{3}$, T. Sato$^{3}$, F. Kagawa$^{1,3}$, R. Kurihara$^{4}$, M. Tokunaga$^{4}$, and Y. Tokura$^{1,3,5}$}

\affiliation{$^{1}$Department of Applied Physics, University of Tokyo, Tokyo 113-8656, Japan \\ 
$^{2}$Division of Materials Science, University of Tsukuba, Tsukuba 305-8573, Japan \\
$^{3}$RIKEN Center for Emergent Matter Science (CEMS), Wako 351-0198, Japan \\
$^{4}$The Institute for Solid State Physics, University of Tokyo, Kashiwa 277-8581, Japan \\
$^{5}$Tokyo College, University of Tokyo, Tokyo 113-8656, Japan
}

\date{\today}

\begin{abstract}
The interplay between electron correlation and topology of relativistic electrons may lead to a new stage of the research on quantum materials and emergent functions. The emergence of various collective electronic orderings/liquids, which are tunable by external stimuli, is a remarkable feature of correlated electron systems, but has rarely been realized in the topological semimetals with high-mobility relativistic electrons. Here, we report that the correlated Dirac electrons with the Mott criticality in perovskite {\CaIrO} show unconventional field-induced successive metal-insulator-metal crossovers in the quantum limit accompanying a giant magnetoresistance (MR) with MR ratio of 3,500 \% (18 T and 1.4 K). The analysis shows that the insulating state originates from the collective electronic ordering such as charge/spin density wave promoted by electron correlation, whereas it turns into the quasi-one-dimensional metal at higher fields due to the field-induced reduction of chemical potential, highlighting the highly field-tunable character of correlated Dirac electrons. 

\end{abstract}

\maketitle

\section{Introduction}
The quantum phenomena of relativistic electron (Dirac/Weyl electron) in solids have been a subject of great interest in modern materials physics. Topological semimetals offer a fertile field of materials to study unique quantum transport phenomena of high-mobility relativistic electrons, presenting the chiral anomaly or various topological phases \cite{2018NPArmitageRevModPhys}. Most of them have been understood in the scheme of single particle physics so far, but there is growing interest in the strong electron correlation as a new route to realize and control the emergent collective topological phenomena \cite{2012HWeiPRL, 2013EGMoonPRL, 2013ZWangPRB, 2014ASekineJPSJ, 2019JGoothNature}. According to the conventional wisdom, the Landau-Fermi liquid picture tends to collapse in the quasi-two/one-dimensional correlated electron system, resulting in or from various charge/spin correlation. In particular, it has been demonstrated that the electronic liquids or charge/spin ordering are often highly sensitive to external stimuli, resulting in the colossal magnetoresistance or pressure induced high-$T_{\rm c}$ superconductivity \cite{2001JFMitchellJPhysChemB, 1996MUeharaJPSJ}. In this context, the spatial confinement of high-mobility relativistic electron with the strong electron correlation may be a promising pathway to find emergent collective topological phenomena. 

In the case of bulk topological semimetals, the quasi-one-dimensional (1D) confinement of relativistic electrons is typically realized in the quantum limit (QL) under a sufficiently strong magnetic field $B$. Electrons in the lowest Landau level with the index $n$=0 are confined in a scale of magnetic length $l_{\rm B} = \sqrt{\hbar⁄eB}$ within a plane perpendicular to the magnetic field, whereas the momentum along the magnetic field is preserved. Previous theoretical studies proposed that various nontrivial phases such as the axion charge density wave (CDW) and excitonic insulator are induced, if the QL of the Dirac/Weyl electron can be realized in materials with the strong electron correlation \cite{2015BRoyPRB, 2016XLiPRB, 2019ZPanPRB}. However, the experimental realization of quantum limit in the strongly correlated electron material is a challenge, and hence the emergence of collective ordering and/or liquid of high-mobility relativistic electron has rarely been demonstrated so far. 

In this context, the correlated Dirac semimetal of perovskite {\CaIrO} provides an ideal arena to study the collective phenomena of high-mobility relativistic electrons in the QL. In {\CaIrO}, due to the strong spin-orbit coupling and electron correlation, the nominally half-filled {\jeff}$=1/2$ band, which lies near the Fermi energy ({\EF}), constitutes the nearly compensated semimetal state with a few electron- and hole-pockets. It has been proposed that the electron pocket emerging around U-point in the Brillouin zone is caused by the Dirac band dispersion with a closed line-node protected by the nonsymmorphic crystalline symmetry ($Pbnm$) as illustrated in the inset to Fig. 1(a) \cite{2012MAZebPRB, 2016YChenPRB, 2015YChenNatCommun}. Recently, it has been shown that the line node is precisely tuned close to {\EF} ($\sim10$ meV below {\EF}) and yields the Dirac electrons with dilute carrier density (less than $2 \times 10^{17}$ {\nUnit}), and high mobility exceeding 60,000 {\muUnit} due to strong electron correlation in the proximity to the Mott criticality \cite{2019JFujiokaNatCommun, 2019MMasukoAPL, 2019RYamadaPRL, 2021JFujiokaPRB}. Specifically, the Fermi velocity is renormalized to a moderately small value ({\vF} $\sim 8 \times 10^{4}$ m/s) compared to other topological materials \cite{2005KSNovoselovNature, 2005YZhangNature, 2010DXQuScience, 2015TLiangNatMater}. Consequently, the correlated Dirac electrons reach the QL at a modest magnetic field less than 10 T. However, the transport property in the QL of this material has not been explored so far, and a possibly striking feature of collective phenomena of the Dirac electrons remains elusive. Here, by magneto-transport measurements and theoretical modeling, we show that the two successive metal-insulator crossovers accompanying a giant magnetoresistance are induced by moderate magnetic field (10-30 T) in the QL region of {\CaIrO}. One is a crossover from semimetallic state to the charge/spin density wave and another is a reentrance to the quasi-1D metallic state of correlated Dirac electrons.

\section{Results}
\subsection{ Magneto-transport properties in the quantum limit.} 
As shown in Fig. 1(a), the resistivity along the $a$-axis, {\Rhozz}, shows a metallic behavior above 150 K, but a peak is observed around 20 K; here we take the Cartesian coordinate with respect to the orthorhombic axes as $x \parallel b, y \parallel c, z \parallel a$ throughout this paper. The peak is attributed to the competition between the reduction of thermally excited carriers and enhancement of carrier mobility \cite{2019JFujiokaNatCommun}. Figure 1(b) shows the low field region of longitudinal magnetoresistivity measured with the electric current and magnetic field parallel to the $a$-axis ({\BaIaz}). With increasing magnetic field, {\Rhozz} initially increases up to 2 T, and then moderately decreases up to 10 T accompanying the Shubnikov-de Haas (SdH) oscillations due to the electron pocket around the line node \cite{2019JFujiokaNatCommun}. The frequency of the SdH oscillations is 10.5 T [see Fig. S1 \cite{Supple}], which roughly corresponds to the carrier density of $2.0 \times 10^{17}$ {\nUnit}, given that the Fermi surface is spherical. Due to the low carrier density, the Dirac electron reaches QL at the modest field of {\BQL} = 6 T, which is defined as the field of SdH oscillation peak due to the Landau index $n$=1 [see Fig. S1 \cite{Supple}]. 

The magnetoresitivity {\Rhozz} ({\BaIa}) up to 55 T is shown in Fig. 1(c). At 1.4 K, {\Rhozz} steeply increases above the threshold magnetic field ({\Bth}) of 10 T, and is nearly saturated around 18 T with the MR-ratio $[\rho_{zz}(B) - \rho_{zz}(0)]/\rho_{zz}(0)$ of 35. Above 18 T, the magnetoresistivity monotonically decreases up to 55 T. With increasing temperature, the peak in {\Rhozz} around 18 T is gradually smeared out, while shifting toward higher magnetic field. On the contrary, the magnetoresistivity in transverse configuration {\Rhoxx} ({\BaIb}) monotonically increases with a small kink at {\Bth} [see Fig. 1(c)].

\begin{figure}
\begin{center}
\includegraphics[width=3.375in,keepaspectratio=true]{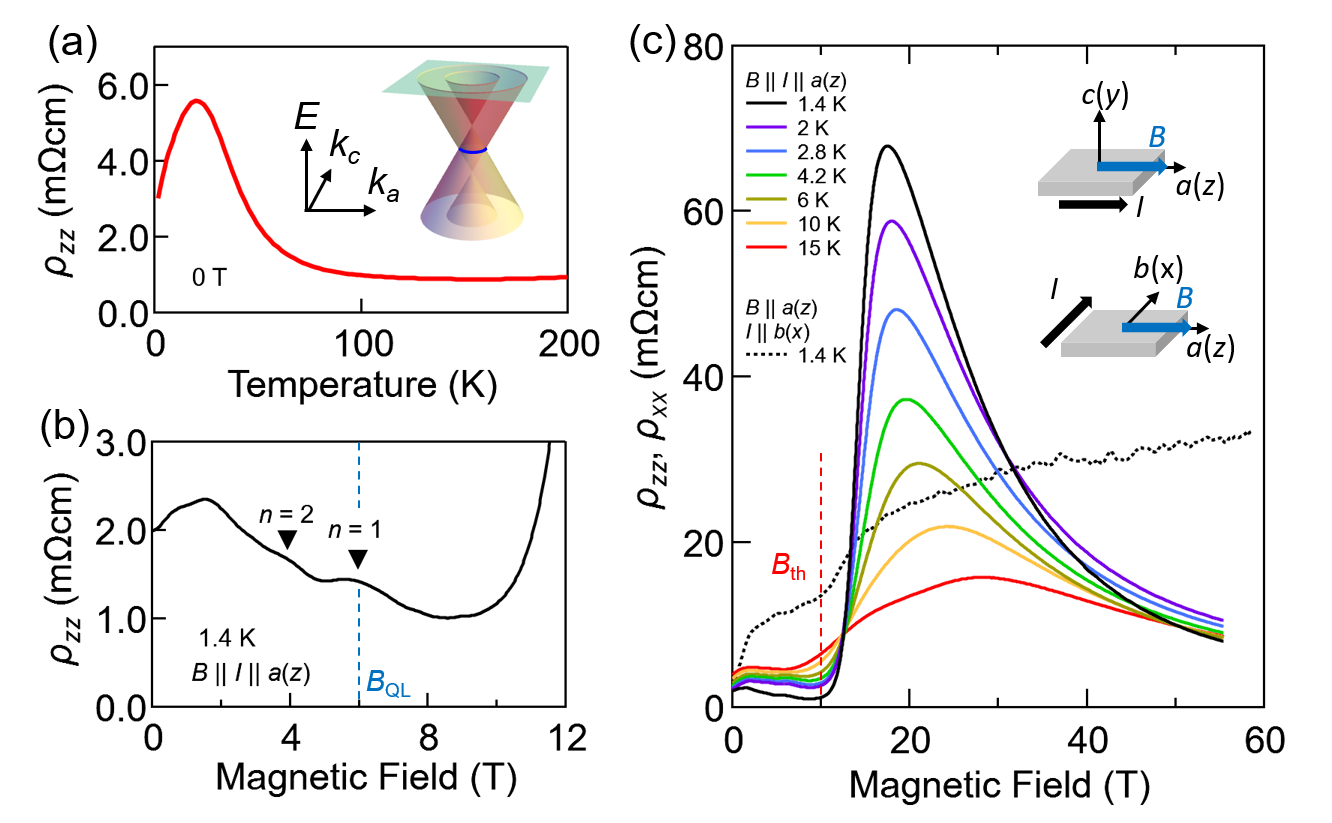}
\caption{ 
(a) Temperature dependence of the resistivity. The inset is a schematic illustration of energy dispersion in $k_{a}$-$k_{c}$ plane near the line node (blue line) near the Fermi energy (pale green plane).
(b) Shubnikov-de-Haas (SdH) oscillations in magnetoresistivity in the longitudinal configuration ({\BaIaz}). {\BQL} and {\Bth} correspond to the magnetic field where quantum limit is reached and where the MR starts to increase steeply.
(c) Magnetoresistivity in the longitudinal configuration ({\BaIaz}) at various temperatures. Magnetoresistivity in the transverse configuration ({\BazIbx}) at 1.4 K is also shown as black dotted line.
}
\end{center}
\end{figure}

To quantify the anisotropy of transport properties in the QL, we compare the magnetoconductivity in the longitudinal configuration ({\BaIa}, {\Gzz}) and that in transverse configuration ({\BaIb}, {\Gxx}) in Fig. 2(a). Here, {\Gzz} is defined as the inverse of {\Rhozz} in the longitudinal configuration ({\BaIaz}) and {\Gxx} as {\Rhoxx}⁄({\Rhoxx}$^2$+{\Rhoyx}$^2$) with {\Rhoxx} and {\Rhoyx} being the resistivity measured in transverse configuration ($B \perp I$; {\BazIbx}) and the Hall resistivity [see Fig. S3 \cite{Supple}], respectively. With increasing the magnetic field, {\Gzz} initially increases, steeply drop around {\Bth}, and shows an upturn at 18 T. Specifically, {\Gzz} at 55 T is enhanced to about ten times compared to that at 18 T. On the contrary, {\Gxx} monotonically decreases with increasing magnetic field with a small drop at {\Bth}. Around {\BQL}, {\Gzz} is about ten times as large as {\Gxx} in agreement with the quasi-one-dimensional confinement of electrons along $B \parallel z$. On the other hand, {\Gzz} falls down to a value comparable to or even less than {\Gxx} around 18 T, but becomes larger than {\Gxx} again in high magnetic field above 30 T. 

Figure 2(b) shows the temperature dependence of {\Gzz} below 15 K. With decreasing temperatures, {\Gzz} monotonically increases at 0 T, but monotonically decreases at 18 T, implying that the ground state changes from a metal to an insulator or a charge-gapped state. On the other hand, the metallic behavior is recovered at 55 T, while {\Gzz} shows a modest temperature dependence. To quantify the insulating behavior at 18 T, we fitted {\Gzz} by the Arrhenius model $\sigma_{zz} = \sigma_{0} + \sigma_{1} \exp(-\Delta_{\rm exp}⁄(2k_{\rm B} T))$ and derived the activation energy {\DeltaExp} with {\Gzero} and {\Gone} being the temperature independent parameters; here $\sigma_{0} (=0.17 \sigma_{1}$) represents the small residual conductivity likely coming from another hole pocket [Supplementary Note 2\cite{Supple}]. The temperature dependence of {\Gzz} below 6 K is well described by the Arrhenius model [see Fig. 2(c)] and {\DeltaExp} is deduced to be 1.2 meV at 18 T. We conducted the similar analyses at various magnetic fields and plotted {\DeltaExp} as a function of magnetic field in Fig. 2(d). The result shows a peak around 18 T in agreement with the behavior of MR at 1.4 K [see Fig. 1(c)]. This suggests that the large peak structure in MR (dip structure of {\Gzz}) originates from the field variation of {\DeltaExp}.

\begin{figure}
\begin{center}
\includegraphics[width=3.375in,keepaspectratio=true]{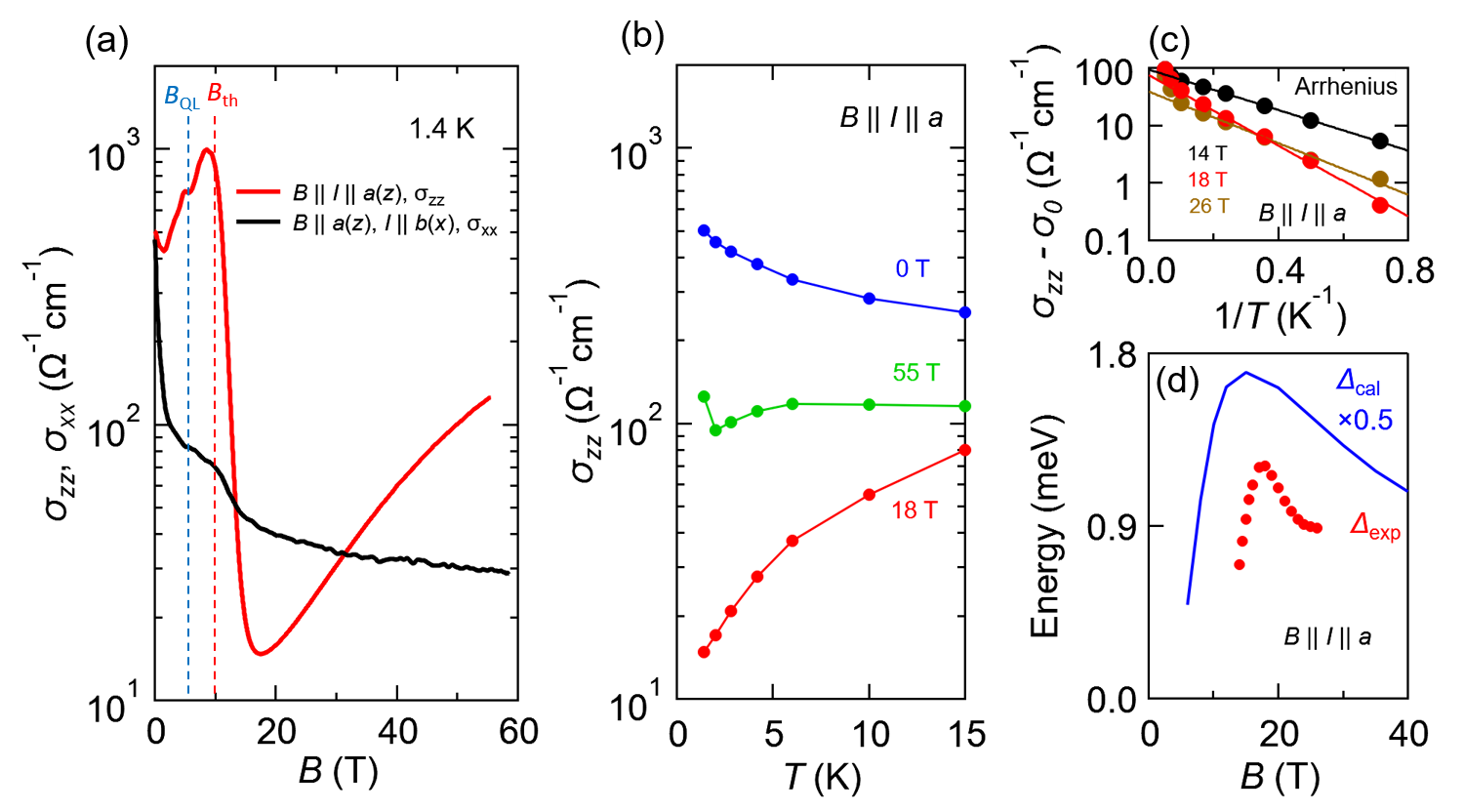}
\caption{
(a) Comparison of the magnetoconductivity in the longitudinal configuration ({\BaIaz}) and that in the transverse configuration ({\BazIbx}). 
(b) Temperature dependence of conductivity at 0, 18, and 55 T. 
(c) Temperature dependence of conductivity plotted as a function of 1/$T$ at 14.5, 17, and 22 T for the Arrhenius model. 
(d) Magnetic field dependence of activation energy extracted by Arrhenius model ({\DeltaExp}) and the gap size of CDW calculated by Fukuyama's model ({\DeltaCal}).
}
\end{center}
\end{figure}

\subsection{Numerical Calculation of Landau levels.}
To get insights into the electronic states under the magnetic field, we numerically calculate the field dependence of the Landau levels (LLs). We construct a tight binding model for the electronic structure around the line node, which takes into account the Zeeman term (g-factor of 2) \cite{2015JWRhimPRB}. Figure 3(a) shows the energy dispersion of band structure around the U-point at 0 T along the $k_{z}$-direction. The Fermi energy is set to be 5.0 meV above the band crossing point, i.e., the line node \cite{2019JFujiokaNatCommun}. Under the magnetic field, the energy band splits into the LLs, and in the QL ($B > ${\BQL} = 6 T), only the $n$=0 LL crosses the Fermi energy [see Figs. 3(b)-(d)]. In the QL, the $n$=0 LL is composed of nearly-degenerate four Weyl bands; they are degenerate at $k_{z}$=0 and split into two doubly degenerate ones at $k_{z} \neq 0$ mainly due to the Zeeman term. Figure 3(e) displays the density of states (DOS) at various magnetic fields. At a small magnetic field (0.1 T), the DOS around {\EF} is small as is expected from the small size of the line node \cite{2019JFujiokaNatCommun}. At 10 T, the DOS in the low energy region (-6 meV $\leq E \leq$ 6 meV), which is governed by the $n$=0 LL, is flat and several peaks due to the $n$$\neq$0 LLs are discernible above 6 meV as well as below -6 meV [see Fig. 3(b)]. At 50 T, the DOS in the region -10 meV $\leq E \leq$ 10 meV is governed by $n$=0 LL and becomes fairly large due to the increased LL degeneracy. As shown in Fig. 3(f), the DOS at {\EF} linearly increases as a function of magnetic field. We note that the charge transfer between these LLs and hole pockets nearby the $\Gamma$-point is negligible and hence the carrier density in the LLs is likely conserved at least below 55 T [Supplementary Note 2 \cite{Supple}]. This suggests that the energy difference ({\DeltaEF}) between the Fermi energy and the band crossing point, as well as the Fermi wave number {\kF}, decreases as a function of magnetic field [see Fig. 3(g)].

\begin{figure}
\begin{center}
\includegraphics[width=3.375in,keepaspectratio=true]{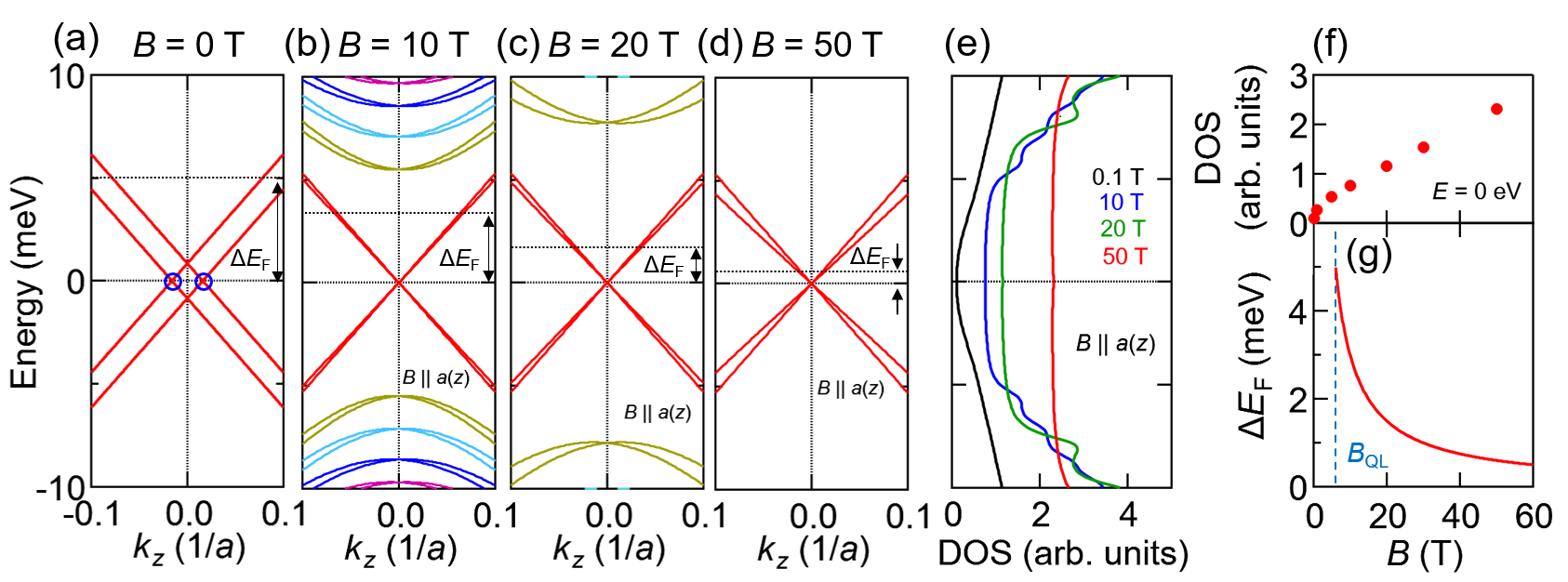}
\caption{
(a) Calculated energy dispersion near the line node. The band crossing due to the line node is denoted by blue circles. Here the origin is U-point in the momentum space [See Fig. S9 \cite{Supple}]. 
(b), (c), and (d) Landau levels around the U-point under the magnetic field of 10, 20, and 50 T, respectively. The magnetic field is applied along a-axis of the crystal ({\Baz}). 
(e) Density of states at $B$ = 0.1, 10, 20, and 50 T. 
(f) The density of states at 0 eV as a function of the magnetic field. 
(g) The magnetic field dependence of {\DeltaEF}. The definition of {\DeltaEF} is shown in (b), (c), and (d).
}
\end{center}
\end{figure}

\subsection{Origin of the non-monotonic field dependence of magnetoresistivity.}
The calculated results suggest that the electronic state of $n$=0 LL is described by the four Weyl bands, which subsist even at sufficiently high magnetic field. This is consistent with the experimental result that the resistivity deep in the QL region at 55 T shows a metallic behavior. In more detail, the conductivity at 55 T shows the power-law type temperature dependence ({\Gzz}$\propto T^{\alpha}, \alpha \sim 0.24$)  in accord with the picture of the Tomonaga-Luttinger liquid [see Fig. S6 \cite{Supple}, \cite{2017XXZhangPRB}]. On the contrary, the insulating behavior with the finite gap around 18 T is far from clear in terms of the numerical calculation. One possible scenario is the magnetic freeze-out, i.e., electron localization by disorders promoted by the quasi-1D confinement. Indeed, the magnetic freeze-out often occurs in the QL of semiconductors such as {\HgCdTe} and InSb \cite{1980JLRobertPhiMagB, 1988MShayeganPRB}, which manifests itself as the non-saturating positive MR. However, this is not the case for the present system wherein the MR shows a peak around 18 T. 

A more plausible scenario would be the field-induced collective electronic ordering promoted by the quasi-one-dimensional confinement. It is known that the Fermi surface of quasi-1D metal is unstable towards a charge-gapped state such as the charge density wave (CDW), spin density wave (SDW) or excitonic insulator \cite{1978HFukuyamaSolStatCommun, 1982YIyePRB, 1984YIyeSolStatCommun, 1999HYaguchiJPSJ, 1998HYaguchiPRL, 1986HeinonenPRB, 1968EWFentonPR}. Although there are several possible candidates of electronic orderings for the quasi-1D state composed of the $n$=0 LL, we focus here on the field induced CDW. By taking into account the long-range Coulomb interaction within the Hartree-Fock approximation, Fukuyama \cite{1978HFukuyamaSolStatCommun} showed theoretically that the electron gas in the QL turns into the charge density wave at low temperatures, as later experimentally observed for graphite \cite{1982YIyePRB, 1984YIyeSolStatCommun, 1999HYaguchiJPSJ, 1998HYaguchiPRL}. We calculated the magnetic field dependence of Tc following Fukuyama's theory \cite{1978HFukuyamaSolStatCommun} and estimated the charge gap {\DeltaCal} (=3.5 $k_{\rm B}T_{\rm c}$) [Supplementary Note 6 \cite{Supple}]. In Fig. 2(d), we show the field dependence of {\DeltaCal} [see also Fig. S10 \cite{Supple}]. {\DeltaCal} shows a peak around 18 T which is consistent with the experimental results within the difference of factor of 3. In more detail, the non-monotonic field dependence of {\DeltaCal} can be understood from the balance between the field variation of the LL degeneracy and {\EF}; the transition temperature, or equivalently the order parameter, initially increases due to the enhancement of the LL degeneracy, but is counterbalanced by the decrease of {\EF} at the higher magnetic field. In this context, the non-monotonic field dependence of magnetoresistivity and {\DeltaExp} can be explained by the field induced CDW formation and its subsequent suppression by higher magnetic fields. We note that the theory considering electron-phonon interaction also gives a qualitatively similar result [Supplementary Note 6 \cite{Supple}, \cite{1986HeinonenPRB}].

\begin{figure}
\begin{center}
\includegraphics[width=3.375in,keepaspectratio=true]{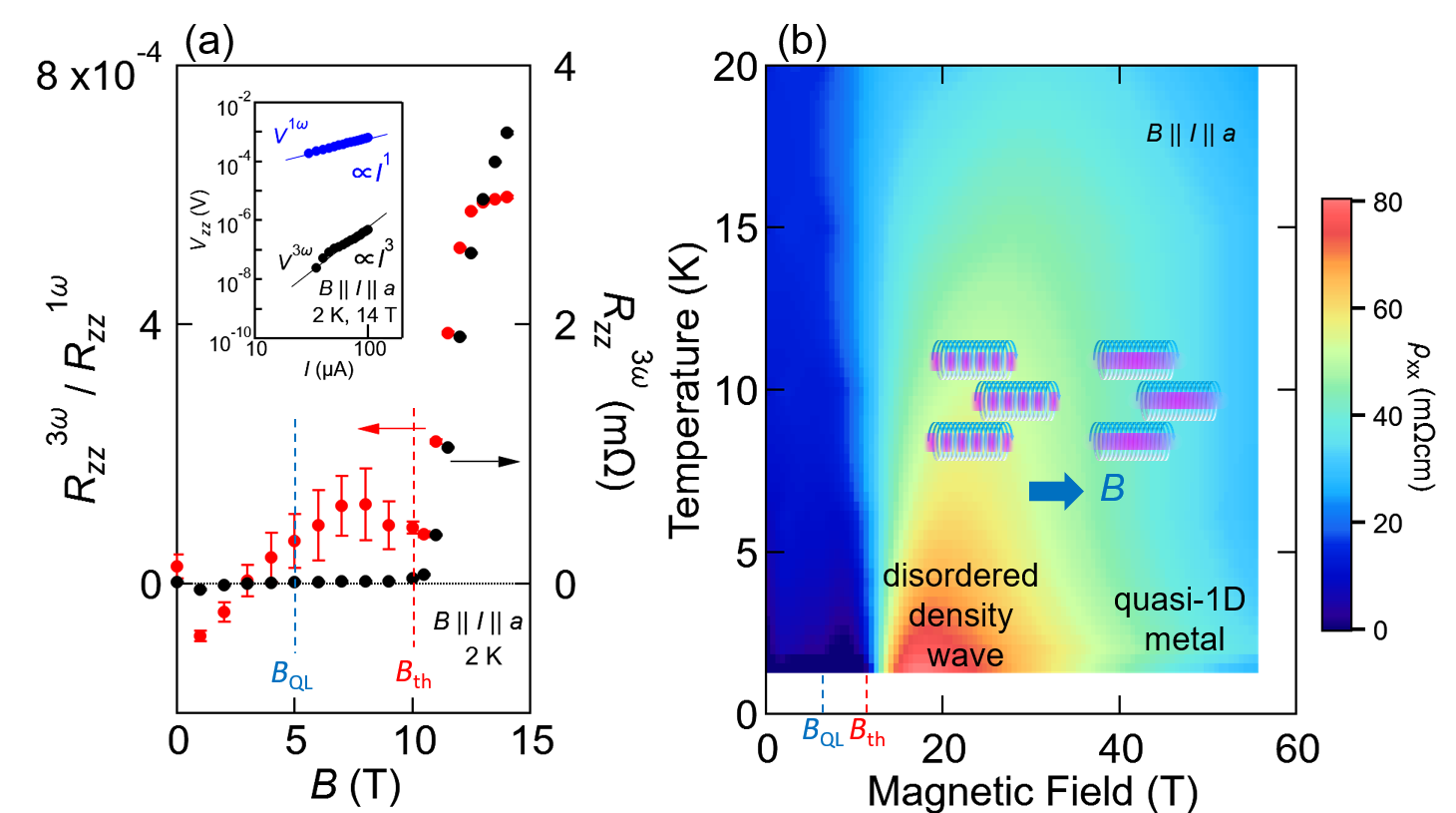}
\caption{
(a) The magnetic field dependence of the third-harmonic resistance 
{\RThrOm}. The third-harmonic resistance normalized by linear resistance {\RThrOmNorm} is also plotted. Here, {\ROneOmDef} and {\RThrOmDef}. The measurements were done under the condition of $I$ = 75 uA and $f$ = 2317 Hz to avoid the heating effects. The inset shows the current dependence of linear and third-harmonic voltage at a frequency of 2317 Hz at 2 K and 14 T. 
(b) The contour plot of resistivity as a function of temperature and magnetic field in the longitudinal configuration ({\BaIaz}). The inset shows schematic illustrations of the disordered density wave phase and quasi-1D metal phase.
}
\end{center}
\end{figure}

\subsection{Nonlinear transport measurements.}
It is instructive to compare these results with the case of graphite \cite{1982YIyePRB, 1984YIyeSolStatCommun}, which is a canonical system showing the field induced insulating state in the QL. In graphite, the insulating state due to 2{\kF} instability is observed as a phase transition accompanying a clear jump of magnetoresistivity at a certain threshold field, which is enhanced at higher temperatures. Moreover, the insulating state returns into a metallic state in the higher magnetic field \cite{1998HYaguchiPRL}. There is a similarity between the both cases, graphite and {\CaIrO}, in terms of overall behavior of meatal-insulator-metal transition (crossover), but the sharpness of the transition is obviously different. To characterize the possible CDW state in {\CaIrO}, we investigated the current-voltage ($I$-$V$) characteristics in the QL. It has been known that the long-range ordered CDW slides over the underlying lattice with a current excitation larger than a certain threshold value, which manifests itself as the non-Ohmic $I$-$V$ characteristics \cite{1988GGrunerRevModPhys}. On the contrary, the disordered or short-range ordered CDW does not show a clear onset of sliding motion. As shown in the inset to Fig. 4(a), even at the sufficiently large current of 100 ${\rm \mu}$A, which corresponds to the electric field about 45 mV/cm, the $I$-$V$ property measured at 2 K and 14 T does not show any clear threshold or jump characteristic of the sliding motion of long-range ordered CDW \cite{1979RMFlemingPRL, 1985YIyePRL}. Alternatively, we found that the $I$-$V$ property already includes the non-Ohmic component even in the weak current region by measurements of third-harmonic voltage response {\VThrOm} [see the inset to Fig. 4(a), Supplementary Note 5 \cite{Supple}]. In Fig. 4(a), we show the third-harmonic resistivity {\RThrOm}, which is defined as $-${\VThrOm}/$I$, as well as that normalized by the linear component of resistivity ({\RThrOmNorm}). Both {\RThrOm} and {\RThrOmNorm} steeply increases above {\Bth}, suggesting that the non-Ohmic behavior is inherent to the QL state. Specifically, the {\RThrOmNorm} reaches $6 \times 10^{-4}$ at 14 T. Such higher harmonic of resistance is often observed in disordered metal or correlated electron systems and typically originates from the spatially inhomogeneous electronic state \cite{1989MADubsonPRB, 2013RRommelPhysSatSolB, 2009VMoshnyagaPRB}. In this context, the non-Ohmic $I$-$V$ property without clear threshold field suggests that the CDW state in the present system is not spatially homogeneous, or equivalently, is not of long-range nature. From the value of 2{\kF}, the period of the CDW is estimated to be about 50 nm at 18 T, and hence the long-ranged CDW formation would require the extremely clean sample, which may not be obtained in the present study.

\begin{table}[h]
	\centering
		\begin{tabular}{|c|c|c|c|}\hline
			\ Material \ & \ {\Bth} (T) \ & \ {\BQL} (T) \ & \ Type of order \ \\ \hline
			{\CaIrO}  & $10$ & $6$ & Disordered density wave \\ \hline
			Graphite  & $25$ & $7$ & density wave \\ \hline
			TaAs      & $80$ & $8$ & Wigner crystallization \\ \hline
			{\ZrTe} & $2$ & $1.3$ & 3D quantum Hall state \\ \hline
		\end{tabular}
	\caption{
	The threshold magnetic field ({\Bth}) necessary to realize the correlated ordered state in quantum limit is compared among the various materials including graphite \cite{1982YIyePRB, 1984YIyeSolStatCommun, 1999HYaguchiJPSJ}, TaAs \cite{2018BJRamshawNatCommun}, and {\ZrTe} \cite{2019FTangNature}. {\CaIrO} enters the correlated ordered state in QL at the remarkably small magnetic field among the reported materials, {\BQL} of which are on the same order of the magnitude main text for more details]. {\ZrTe} shows a remarkably small {\Bth}, but it would possibly be due to the small {\BQL} of the material.
	}
	\label{Table_Comparison}
\end{table}

\section{Discussion}
On the basis of these results, we constructed the electronic phase diagram as shown in Fig. 4(b). In the low-field regime below {\Bth}, the metallic state extends over a wide temperature range. In the intermediate field regime above {\Bth}, the insulating state as characterized by the disordered CDW emerges at low temperatures, but gradually turns into the quasi-1D metallic state perhaps with the nature of Tomonaga-Luttinger liquid in a higher field regime. Although we focus on the CDW model here, the possibility of other electronic orderings, such as SDW or excitonic insulator cannot be excluded due to the multi-band structure of $n$=0 LLs and the strong electron correlation. However, we anticipate that the CDW picture gives a good starting point to capture the field induced insulating state in the present system. Moreover, we note here that the threshold magnetic field of the insulating state ({\Bth}) is as small as 10 T. Except the case of {\ZrTe} with exceptionally small {\BQL} ($\sim$2 T) \cite{2019FTangNature}, this value is much smaller than the value of other semimetals, such as graphite \cite{1982YIyePRB, 1984YIyeSolStatCommun, 1999HYaguchiJPSJ, 1998HYaguchiPRL} ({\Bth}=25T) and TaAs ({\Bth}=80T) \cite{2018BJRamshawNatCommun}, regardless of similar value of the {\BQL} [see Table 1]. The theoretical model of CDW as presented above also demonstrates that {\Bth} decreases as {\vF} decreases through the enhancement of density of state, when {\vF} is in the range of $10^4$-$10^6$ m/s [see Fig. S11 and Supplementary Note 6 \cite{Supple}]. In this context, the moderate renormalization of {\vF} due to the strong electron correlation \cite{2019JFujiokaNatCommun} may be one of the key ingredients for the small value of {\Bth} in {\CaIrO}.

\section{Conclusion}
In this study, we have investigated the quantum limit (QL) transport of correlated Dirac electrons in the perovskite {\CaIrO} by means of magneto-transport measurements and theoretical calculations. In the QL, the magnetoresistivity steeply increases around 10 T (= {\Bth}) and the insulating state with a finite energy gap emerges around 18 T, resulting in the giant magnetoresistance ratio of 3,500 \%. By further increasing the magnetic field, both the gap and resistivity dramatically decrease, resulting in the quasi-1D metallic state in the deep QL regime. The non-monotonic field dependence of the gap as well as the non-Ohmic current-voltage characteristics suggests that the field-induced insulating state originates from the collective electronic ordering, likely the charge density wave, spin density wave or excitonic insulator driven by the Fermi surface instability inherent to the quasi-one-dimensional $n$=0 Landau levels. The field-induced crossover between the metallic state and the gapped state occurs in the fairly low magnetic field regime (10-30 T) among the conventional semimetals, highlighting the highly field-sensitive character of strongly correlated Dirac electrons relevant to the Mott criticality.

\section{Methods}
Single crystalline sample of perovskite {\CaIrO} were synthesized under high pressures using the cubic-anvil type facility. The samples were treated at pressure of 1 GPa and temperature of 1200 ${}^\circ\mathrm{C}$. The materials were kept under this condition for 10 min and then quenched to room temperature. The typical size of the sample is about 0.5×0.3×0.3 mm$^{3}$. We have determined the crystal orientation by an in-house X-ray diffractometer. The crystal structure is identified as the orthorhombic ({\GdFeO}-type) perovskite with the space group of $Pbnm$.

The resistivity ({\Rhozz}) in longitudinal configuration and the resistivity ({\Rhoxx}) and Hall resistivity ({\Rhoyx}) in transverse configuration were measured by a four- or five- probe method with indium electrode. Epo-tek H20S silver epoxy and 50-${\rm \mu m}$-diameter gold wires were used to form electrical contacts. The magnetic field was applied along the a-axis and the current was applied along the $a$-axis and $b$-axis of the crystal for a longitudinal and transverse configuration, respectively. The magnetotransport measurements up to 55 T in a temperature range from 1.4 to 40 K were done under pulsed high magnetic fields using nondestructive magnets installed at The Institute for Solid State Physics, The University of Tokyo. The magnetic fields up to 55 T (pulse durations of 36 ms) were generated using bipolar pulse magnets, which can generate both positive and negative fields. We used numerical lock-in technique at a frequency of 100 kHz. {\Rhozz} and {\Rhoxx} are symmetrized and {\Rhoyx} is antisymmetrized with respect to the magnetic field.

Measurements of the third-harmonic voltage response have been performed by standard four-terminal geometry in the longitudinal configuration ({\BaIaz}). An ac current $I=I_{0}  \sin \omega t $ is applied to the sample and the linear ($V^{1\omega}$) and nonlinear response ($V^{3\omega}$) in the total voltage signal $V = V^{1\omega} \sin \omega t + V^{3\omega} \sin 3\omega t + \cdots$ are detected simultaneously by two digital lock-in amplifiers. The magnetic fields up to 14 T is applied and the sample was cooled down to the temperature of 2 K using the Physical Property Measurement System (Quantum Design). The measurements were done in a frequency range from 3 to 4,000 kHz.

To calculate the Landau levels (LLs) of line node, we constructed a low-energy effective model of the band structure around the U point, where the line node resides, based on the previous works \cite{2016YChenPRB, 2015JWRhimPRB} on {\SrIrO}, which has the same crystal symmetry as the perovskite {\CaIrO} [Supplementary Note 7 \cite{Supple}]. Using the Landau quantization for {\Bparaa}, we derived coupled secular equations similar to those derived in Ref. \cite{2015JWRhimPRB} for {\Bparac}. We then numerically solved these equations, to obtain the LLs. The DOS is calculated by integrating the weight ($\propto B$) of the LLs over $-0.48/a < k_{z} < 0.48/a$ around the U point.

\section{Acknowledgement}
Authors thank A. Tsukazaki, M. Kawasaki, N. Nagaosa, A Yamamoto, Y. Kaneko, S. Ishiwata, Y. Fuseya, Y. Awashima, M. Masuko, and R. Kaneko for fruitful discussion. This work was supported by a Japan Society for the Promotion of Science KAKENHI (Grants No. 20J21312, No. 16H00981, No. 18H01171, No. 18H04214, and No. 16H06345) from the MEXT and by PRESTO (Grant No. JPMJPR15R5) and CREST (Grants No. JPMJCR16F1), Japan Science and Technology Japan.



\begin{thebibliography}{99}

\bibitem{2018NPArmitageRevModPhys} N. P. Armitage, E. J. Mele, A. Vishwanath, Weyl and Dirac semimetals in three-dimensional solids. Rev. Mod. Phys. {\bf 90}, 015001 (2018).
\bibitem{2012HWeiPRL} H. Wei, S. -P. Chao, V. Aji, Excitonic phases from Weyl semimetals. Phys. Rev. Lett. {\bf 109}, 196403 (2012).
\bibitem{2013EGMoonPRL} E. -G. Moon, C. Xu, Y. B. Kim, L. Balents, Non-Fermi-liquid and topological states with strong spin-orbit coupling. Phys. Rev. Lett. {\bf 111}, 206401 (2013).
\bibitem{2013ZWangPRB} Z. Wang, S. -C. Zhang, Chiral anomaly, charge density waves, and axion strings from Weyl semimetals. Phys. Rev. B {\bf 87}, 161107(R) (2013).
\bibitem{2014ASekineJPSJ} A. Sekine, K. Nomura, Axionic antiferromagnetic insulator phase in a correlated and spin-orbit coupled system. J. Phys. Soc. Jpn. {\bf 83}, 104709 (2014).
\bibitem{2019JGoothNature} J. Gooth, B. Bradlyn, S. Honnali, C. Schindler, N. Kumar, J. Noky, Y. Qi, C. Shekhar, Y. Sun, Z. Wang, B. A. Bernevig, C. Felser, Axionic charge-density wave in the Weyl semimetal (TaSe$_4$)$_2$I. Nature {\bf 575}, 315-319 (2019).
\bibitem{2001JFMitchellJPhysChemB} J. F. Mitchell, D. N. Argyriou, A. Berger, K. E. Gray, R. Osborn, U. Welp, Spin, Charge, and Lattice States in Layered Magnetoresistive Oxides. J. Phys. Chem. B {\bf 105}, 10731-10745 (2001).
\bibitem{1996MUeharaJPSJ} M. Uehara, T. Nagata, J. Akimitsu, H. Takahashi, N. M\^{o}ri, K. Kinoshita, Superconductivity in the Ladder Material Sr$_{0.4}$Ca$_{13.6}$Cu$_{24}$O$_{41.84}$. J. Phys. Soc. Jpn. {\bf 65}, 2764-2767 (1996).
\bibitem{2015BRoyPRB} B. Roy, J. D. Sau, Magnetic catalysis and axionic charge density wave in Weyl semimetals. Phys. Rev. B {\bf 92}, 125141 (2015).
\bibitem{2016XLiPRB} X. Li, B. Roy, S. D. Sarma, Weyl fermions with arbitrary monopoles in magnetic fields: Landau levels, longitudinal magnetotransport, and density-wave ordering. Phys. Rev. B {\bf 94}, 195144 (2016).
\bibitem{2019ZPanPRB} Z. Pan, R. Shindou, Ground-state atlas of a three-dimensional semimetal in the quantum limit. Phys. Rev. B {\bf 100}, 165124 (2019).
\bibitem{2012MAZebPRB} M. A. Zeb, H. -Y. Kee, Interplay between spin-orbit coupling and Hubbard interaction in {\SrIrO} and related $Pbnm$ perovskite oxides. Phys. Rev. B {\bf 86}, 085149 (2012).
\bibitem{2016YChenPRB} Y. Chen, H. S. Kim, H. -Y. Kee, Topological crystalline semimetals in nonsymmorphic lattices. Phys. Rev. B {\bf 93}, 155140 (2016).
\bibitem{2015YChenNatCommun} Y. Chen, Y. -M. Lu, H. -Y. Kee, Topological crystalline semimetals in orthorhombic perovskite iridates. Nat. Commun. {\bf 6}, 6593 (2015).
\bibitem{2019JFujiokaNatCommun} J. Fujioka, R. Yamada, M. Kawamura, S. Sakai, M. Hirayama, R. Arita, T. Okawa, D. Hashizume, M. Hoshino, Y. Tokura, Strong-correlation induced high-mobility electrons in Dirac semimetal of perovskite oxide. Nat. Commun. {\bf 10}, 362 (2019).
\bibitem{2019MMasukoAPL} M. Masuko, J. Fujioka, M. Nakamura, M. Kawasaki, Y. Tokura, Strain-engineering of charge transport in the correlated Dirac semimetal of perovskite {\CaIrO} thin films. APL Mater. {\bf 7}, 081115 (2019).
\bibitem{2019RYamadaPRL} R. Yamada, J. Fujioka, M. Kawamura, S. Sakai, M. Hirayama, R. Arita, T. Okawa, D. Hashizume, M. Hoshino, Y. Tokura, Large variation of Dirac semimetal state in perovskite {\CaIrO} with pressure-tuning of electron correlation. Phys. Rev. Lett. {\bf 123}, 216601 (2019).
\bibitem{2021JFujiokaPRB} J. Fujioka, R. Yamada, T. Okawa, Y. Tokura, Dirac polaron dynamics in the correlated semimetal of perovskite {\CaIrO}. Phys. Rev. B {\bf 103}, L041109 (2021).
\bibitem{2005KSNovoselovNature} K. S. Novoselov, A. K. Geim, S. V. Morozov, D. Jiang, M. I. Katsnelson, I. V. Grigorieva, S. V. Dubonos, A. A. Firsov, Two-dimensional gas of massless Dirac fermions in graphene. Nature {\bf 438}, 197-200 (2005).
\bibitem{2005YZhangNature} Y. Zhang, Y. -W. Tan, H. L. Stormer, P. Kim, Experimental observation of the quantum Hall effect and Berry's phase in graphene. Nature {\bf 438}, 201-204 (2005).
\bibitem{2010DXQuScience} D. -X. Qu, Y. S. Hor, J. Xiong, R. J. Cava, N. P. Ong, Quantum oscillations and Hall anomaly of surface states in the topological insulator Bi$_2$Te$_3$. Science {\bf 329}, 821-824 (2010).
\bibitem{2015TLiangNatMater} T. Liang, Q. Gibson, M. N. Ali, M. Liu, R. J. Cava, N. P. Ong, Ultrahigh mobility and giant magnetoresistance in the Dirac semimetal {\CdAs}. Nat. Mater. {\bf 14}, 280-284 (2015).
\bibitem{Supple} See Supplemental Material, which includes Refs. \cite{2019JFujiokaNatCommun, 2001CBiaginiEPL, 2017XXZhangPRB, 1989MADubsonPRB, 1992YYagilPRB, 1978HFukuyamaSolStatCommun, 2021JFujiokaPRB, 1986HeinonenPRB, 2015JWRhimPRB, 2015YChenNatCommun}, for additional information on the transport measurements, analyses of quantum oscillation, and the numerical calculation on Landau levels.
\bibitem{2015JWRhimPRB} J. -W. Rhim, Y. B. Kim, Landau level quantization and almost flat modes in three-dimensional semimetals with nodal ring spectra. Phys. Rev. B {\bf 92}, 045126 (2015). 
\bibitem{2017XXZhangPRB} X. -X. Zhang, N. Nagaosa, Tomonaga-Luttinger liquid and localization in Weyl semimetals. Phys. Rev. B {\bf 95}, 205143 (2017). 
\bibitem{1980JLRobertPhiMagB} J. L. Robert, A. Raymond, R. L. Aulombard, C. Bousquet, Magnetic freeze-out in doped semiconductors the metal non-metal transition in $n$-type InSb. Philos. Mag. B {\bf 42}, 1003-1025 (1980).
\bibitem{1988MShayeganPRB} M. Shayegan, V. J. Goldman, H. D. Drew, Magnetic-field-induced localization in narrow-gap semiconductors {\HgCdTe} and InSb. Phys. Rev. B {\bf 38}, 5585-5602 (1988).
\bibitem{1978HFukuyamaSolStatCommun} H. Fukuyama, CDW instability of electron gas in a strong magnetic field. Solid State Commun. {\bf 26}, 783-786 (1978).
\bibitem{1982YIyePRB} Y. Iye, P. M. Tedrow, G. Timp, M. Shayegan, M. S. Dresselhaus, G. Dresselhaus, A. Furukawa, S. Tanuma, High-magnetic-field electronic phase transition in graphite observed by magnetoresistance anomaly. Phys. Rev. B {\bf 25}, 5478-5485 (1982).
\bibitem{1984YIyeSolStatCommun} Y. Iye, P. M. Beglund, L. E. McNeil, The magnetic field dependence of the critical temperature for the electronic phase transition in graphite in the quantum limit. Solid State Commun. {\bf 52}, 975-980 (1984).
\bibitem{1999HYaguchiJPSJ} H. Yaguchi, T. Takamasu, Y. Iye, N. Miura, Non-ohmic out-of-plane transport in a high-magnetic-field-induced phase of graphite. J. Phys. Soc. Jpn. {\bf 68}, 181-184 (1999).
\bibitem{1998HYaguchiPRL} H. Yaguchi, J. Singleton, Destruction of the Field-Induced Density-Wave State in Graphite by Large Magnetic Fields. Phys. Rev. Lett. {\bf 81}, 5193-5196 (1998).
\bibitem{1986HeinonenPRB} O. Heinonen, A. A. -J. Radi, Electron-phonon interactions and charge-density-wave formations in strong magnetic fields. Phys. Rev. B {\bf 33}, 5461-5464 (1986).
\bibitem{1968EWFentonPR} E. W. Fenton, Excitonic insulator in a magnetic field. Phys. Rev. {\bf 170}, 816-821 (1968).
\bibitem{1988GGrunerRevModPhys} G. Gr\"{u}ner, The dynamics of charge-density waves. Rev. Mod. Phys. {\bf 60}, 1129 (1988).
\bibitem{1979RMFlemingPRL} R. M. Fleming, C. C. Grimes, Sliding-mode conductivity in NbSe$_3$: Observation of a threshold electric field and conduction noise. Phys. Rev. Lett. {\bf 42}, 1423-1426 (1979).
\bibitem{1985YIyePRL} Y. Iye, G. Dresselhaus, Non-ohmic transport in the magnetic-field-induced charge-density-wave phase of graphite. Phys. Rev. Lett. {\bf 54}, 1182-1184 (1985).
\bibitem{1989MADubsonPRB} M. A. Dubson, Y. C. Hui, M. B. Weissman, J. C. Garland, Measurement of the fourth moment of the current distribution in two-dimensional random resistor networks. Phys. Rev. B {\bf 39}, 6807-6815 (1989).
\bibitem{2013RRommelPhysSatSolB} R. Rommel, B. Hartmann, J. Brandenburg, J. A. Schlueter, J. M\"{u}ller, Nonlinear electronic transport in the anomalous metallic state of quasi-2D organic superconductors $\kappa$-(BEDT-TTF)$_2$X. Phys. Status Solidi B {\bf 250}, 568-574 (2013).
\bibitem{2009VMoshnyagaPRB} V. Moshnyaga, K. Gehrke, O. I. Lebedev, L. Sudheendra, A. Belenchuk, S. Raabe, O. Shapoval, J. Verbeeck, G. Van Tendeloo, K. Samwer, Electrical nonlinearity in colossal magnetoresistance manganite films: Relevance of correlated polarons. Phys. Rev. B {\bf 79}, 134413 (2009). 
\bibitem{2019FTangNature} F. Tang, Y. Ren, P. Wang, R. Zhong, J. Schneeloch, S. A. Yang, K. Yang, P. A. Lee, G. Gu, Z. Qiao, L. Zhang, Three-dimensional quantum Hall effect and metal–insulator transition in {\ZrTe}. Nature 569, 537-541 (2019).
\bibitem{2018BJRamshawNatCommun} B. J. Ramshaw, K. A. Modic, A. Shekhter, Y. Zhang, E. -A. Kim, P. J. W. Moll, M. D. Bachmann, M. K. Chan, J. B. Betts, F. Balakirev, A. Migliori, N. J. Ghimire, E. D. Bauer, F. Ronning, R. D. McDonald, Quantum limit transport and destruction of the Weyl nodes in TaAs. Nat. Commun. {\bf 9}, 2217 (2018).
\bibitem{2001CBiaginiEPL} C. Biagini, D. L. Maslov, M. Y. Reizer, L. I. Glazman, Magnetic-field-induced Luttinger liquid. Europhys. Lett. {\bf 55}, 383-389 (2001).
\bibitem{1992YYagilPRB} Y. Yagil, G. Deutscher, Third-harmonic generation in semicontinuous metal films. Phys. Rev. B {\bf 46}, 16115 (1992).


\end{thebibliography}
\end{document}